# Exploring water's no-man's land

Peter Lunkenheimer*, Daniel Reuter, Arthur Schulz, Martin Wolf, and Alois Loidl

Experimental Physics V, Center for Electronic Correlations and Magnetism, University of Augsburg, 86159 Augsburg, Germany

The investigation of water's glass transition and a possible liquid-liquid transition within its supercooled state is hampered by its inevitable crystallization in a temperature range, termed "no-man's land". Here we report dielectric-spectroscopy and calorimetry measurements of pure water and various aqueous LiCl solutions, part of the latter being quenched to avoid crystallization. By investigating solutions of relatively low salt content and by covering an exceptionally broad frequency range up to THz, we find strong hints at a crossover in water from a strong to a fragile liquid, characterized by different glass-transition temperatures and different non-Arrhenius temperature dependences of the molecular dynamics.

## I. INTRODUCTION

All liquids can be supercooled into an amorphous glass state, but for some of them the cooling rates required to avoid crystallization are extremely high. This applies for water, which (under ambient pressure) only becomes a glass when deposited as vapor or aerosol on cooled substrates [1,2]. Unfortunately, gathering information on water's glass-liquid transition by heating up such quenched samples is hampered by inevitable crystallization occurring above ~150 K, the lower bound of what is called water's "no-man's land" (NML) [3]. Its upper bound is ~235 K, below which water cannot be supercooled with moderate rates. Therefore, water's glass-transition temperature $T_g$ is disputed, just as the controversial assumption [4,5] of a liquid-liquid transition from a fragile to a strong state, hidden within the NML [1,2,6,7,8,9,10,11,12,13,14]. So-called strong glass formers exhibit only weak deviations of their structural dynamics (quantified by the relaxation time $\tau$) from simple thermally activated temperature-dependence [15]. In fragile ones, these deviations are large, which is commonly ascribed to cooperativity of molecular motions [16].

Investigating supercooled water is highly relevant because many peculiarities of liquid water may be traced back to the proposed fragile-strong transition (FST) [2,3,17]. Moreover, glassy water probably is quite abundant in space [18,19]. To circumvent crystallization in the NML, water with admixed salts or other compounds [6,8,17,20,21,22,23,24,25,26,27], confined water [28,29,30,31], or bound water, e.g., in proteins [32,33,34] was investigated, but one may question the relevance of such results for pure bulk water. While several computer experiments suggest two liquid forms of water [3,35,36,37], even the bare existence of fragile water and a FST is still disputed [7,14,33,38,39,40,41,42].

Here, we investigate pure water and aqueous LiCl solutions, guided by comments from Angell [6,22] that, for low salt contents, indeed an FST should be directly observable, which is relevant for pure water, too. By performing extreme broadband dielectric measurements up to THz frequencies, complemented by DSC experiments, and by supercooling low-concentration samples via quenching, we shed new light on the FST in pure water.

## II. EXPERIMENTAL DETAILS

### A. Samples

The samples were prepared by diluting an aqueous 8 mol/l LiCl solution (Sigma Aldrich) with deionized $H_2O$ (Merck "Ultrapure"). Higher concentrated solutions were made by mixing deionized $H_2O$ with pure LiCl salt (ChemPur), which was stored and weighed in argon atmosphere to prevent water absorption. Concentrations are given in mol%. Please note that in a previous publication [26] we specified the number (percentage) of LiCl molecules per one water molecule. In the present work, we use mol% LiCl to be compatible with published work treating concentration-dependent data. By dielectric spectroscopy, we investigated solutions with LiCl concentrations of 1.8, 3.6, 5.0, 7.3, and 14.8 mol%, which corresponds to molarities of 1, 2, 2.75, 4, and 8 mol/l, respectively.

### B. Dielectric measurements

For dielectric spectroscopy, a combination of several techniques was used: At frequencies up to 1 MHz, we employed a frequency-response analyzer (Novocontrol Alpha-A). In the interval from 1 MHz to 3 GHz, a coaxial reflectometric setup including impedance analyzers (Agilent 4294A and Agilent E4991A) was used. In both cases, the samples were put into parallel-plate capacitors. The frequency range from 100 MHz to 40 GHz was covered by a coaxial open-end reflection technique, using the Agilent "Dielectric Probe Kit" and the Agilent E8363B Network Analyzer. For the highest frequencies up to about 1 THz, a terahertz time-domain spectrometer TPS Spectra 3000 by Teraview was used. More information about the measurement principles and setups can be found in [26] and references therein.

*Contact author:
peter.lunkenheimer@physik.uni-augsburg.de

For cooling, a $N_2$ gas cryostat (Novocontrol Quatro) with home-made insets, a helium-flow cryostat, and Peltier elements were employed. For quenching, performed in the gas cryostat at frequencies up to 3 MHz, the complete cryostat inset, including the wired and filled sample capacitor, was first immersed into liquid nitrogen before putting it into the precooled cryostat. The achieved cooling rate was too fast to be properly monitored. In earlier DSC measurements [20,21,27], aqueous LiCl solutions with $x < 9$ mol% were found to be impossible to supercool by liquid-nitrogen quenching. Probably the achieved cooling rates were higher and/or the sample environment (capacitors with polished stainless-steel plates in $N_2$ atmosphere) favored the avoidance of crystallization in the present experiments.

### C. Differential scanning calorimetry

DSC measurements were performed with a DSC 8500 (Perkin Elmer). The device was calibrated for heating measurement runs with a 3-point method using n-dodecane, n-heptane, and indium as standard samples. For the measurements, small amounts of sample material (< 20 mg) were hermetically sealed into aluminum pans. To determine the fragility index $m$, the measurement procedure was adapted from Wang *et al.* [43]. The "standard scan rate" was set to 10 K/min and the cooling rate was varied between 0.5 and 100 K/min.

## III. EVALUATION OF THE DIELECTRIC SPECTRA

As a typical example, Fig. 1 shows the spectra of the dielectric constant $\varepsilon'$ and loss $\varepsilon''$ as measured at frequencies up to 10 MHz and various temperatures for the quenched 5 mol% sample. To describe the contributions of the dipolar-relaxation process of the water molecules to the measured dielectric spectra, the empirical Cole-Davidson (CD) function [44], often employed to fit dielectric spectra of dipolar glass formers [45,46] and also of pure water [26], was used. Via the common relation $\varepsilon'' \propto \sigma'/\nu$ (where $\sigma'$ is the real part of the conductivity), the ionic charge transport in aqueous LiCl solutions leads to an inevitable additional contribution to the loss spectra arising from the dc-conductivity $\sigma_{dc}$ and dominating at low frequencies. In Fig. 1(b), it is revealed by the $1/\nu$ upturn of $\varepsilon''(\nu)$ at low frequencies (not completely shown). At high temperatures, it essentially completely superimposes the expected relaxation peaks related to the reorientational motions of the water molecules while, at low temperatures, these peaks are still evidenced by shoulders in the spectra. In contrast, in $\varepsilon'(\nu)$ [Fig. 1(a)], where there is no contribution from dc charge transport, the sigmoidal steps expected for a dipolar relaxation process are clearly seen.

To better illustrate the molecular relaxation behavior in the loss spectra, the dc contribution can be subtracted which reveals the expected relaxation peaks, from which the relaxation time can be estimated via $\tau \approx 1/(2\pi\nu_{peak})$ (see Fig. 2 for examples). However, to avoid ambiguities in the proper determination of the subtracted $\sigma_{dc}$ values, for all LiCl concentrations (except for 14.8 mol%; see below) we have instead fitted the raw dielectric spectra with the inclusion of a contribution $\varepsilon''_{dc} = \sigma_{dc}/(2\pi\nu\varepsilon_0)$ in the overall fit function ($\varepsilon_0$ denotes the permittivity of vacuum). Moreover, we have also simultaneously fitted the independently measured real part of the permittivity, $\varepsilon'$, which is not hampered by the dc conductivity.

The additional increase of $\varepsilon'(\nu)$ with decreasing frequency, seen at low frequencies in Fig. 1(a), can be ascribed to electrode polarization. This is a well-known effect for ionic conductors and caused by the accumulation of the ions at the sample-electrode interfaces at low frequencies [47]. In the performed fits, it was taken into account by assuming a parallel RC circuit connected in series to the sample, as treated in detail in Refs. [47] and [48]. Finally, at low temperatures and high frequencies, an excess-wing like contribution [45,46,49] is revealed in the loss spectra of Fig. 1(b). It was formally taken into account by up to two Cole-Cole functions [50] as often used for secondary relaxation processes [46,51].

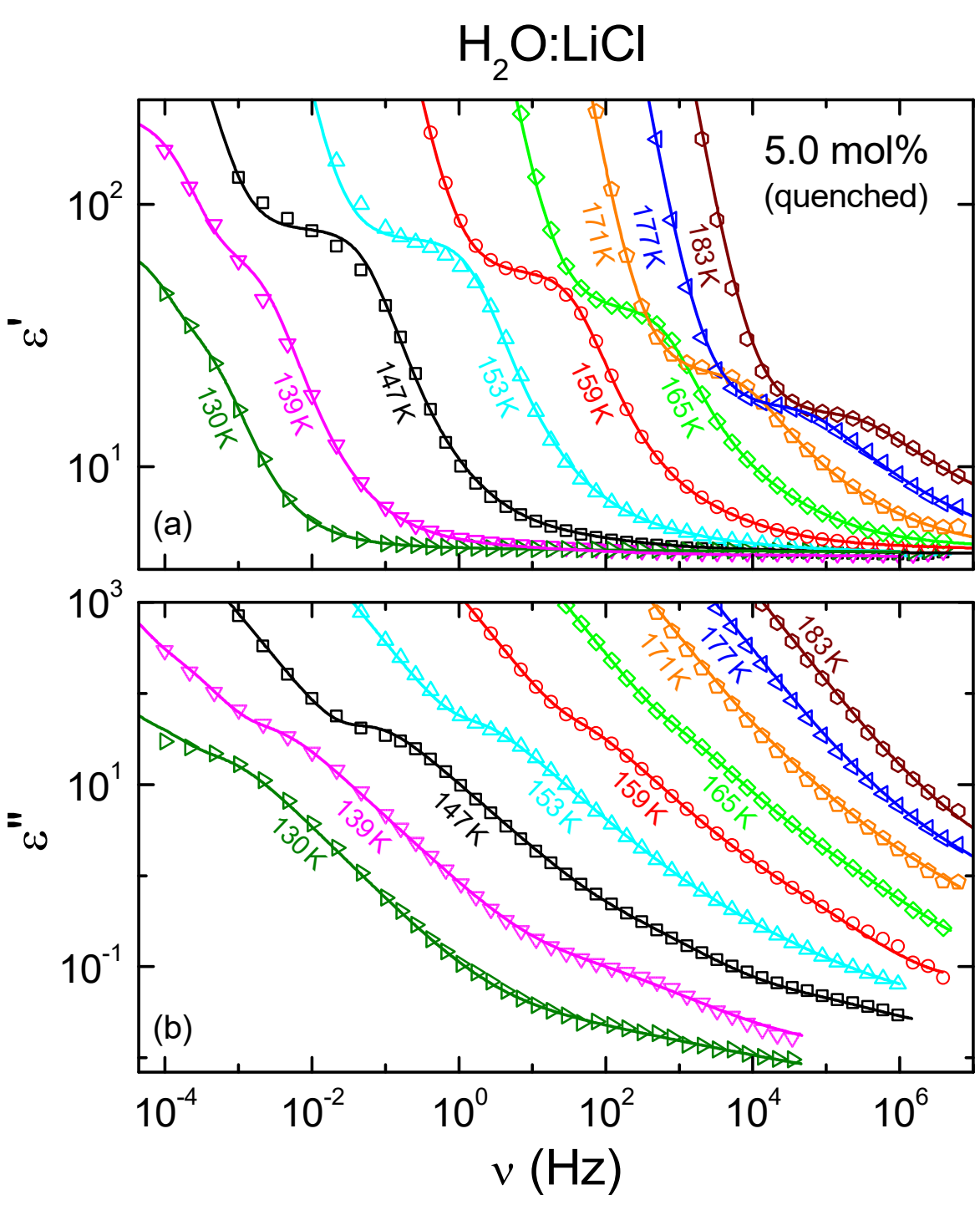

FIG. 1. Complex permittivity spectra of a quenched aqueous solution with 5.0 mol% LiCl. (a) Real part of the permittivity $\varepsilon'$ (dielectric constant). (b) Imaginary part $\varepsilon''$ (dielectric loss). For both quantities, raw data, including conductivity contributions and effects of electrode polarization, are shown on double logarithmic scales for various temperatures. The lines indicate simultaneous fits of $\varepsilon'$ and $\varepsilon''$ as discussed in the text.

As shown by the lines in Fig. 1, reasonable fits of the experimental data could be achieved in this way. It should be noted that for none of the measured spectra, all the different contributions, discussed above, had to be simultaneously used in the fits. For example, at high temperatures the excess-wing like feature is not seen and at low temperatures electrode effects are absent. We also want to point out that, at all temperatures, a step and corresponding point of inflection is detected in the $\varepsilon'$ spectra which clearly restricts the uncertainty of the water relaxation-time obtained from the fits. This is also the case for the results on the other solutions.

An exception of the above-described evaluation method are the data for 14.8 mol% LiCl from Ref. [26] [Fig. 2(a)], where the loss peaks after dc subtraction were reported and fitted by the empirical Havriliak-Negami (HN) function [52]. For this data set, we determined the peak frequencies from the HN fit parameters [53], from which we calculated the relaxation times (see remark in the next paragraph). To check for possible errors in $\tau$ due to this different evaluation procedure, we also analyzed the raw $\varepsilon'$ and $\varepsilon''$ spectra of the 14.8 mol% solution at four temperatures (140, 145, 150, and 155 K) using the same procedure as for the other concentrations. The resulting $\tau$ values did not reveal any significant deviations from those derived from the peak frequencies.

Relaxation peaks in dielectric loss spectra of dipolar liquids are usually broader than expected by the Debye theory, which nowadays is commonly ascribed to a distribution of relaxation times due to heterogeneity [54,55]. As discussed in Ref. [26], even for pure water small deviations from Debye behavior show up. Thus, it is advisable to consider average relaxation times $\langle\tau\rangle$ instead of the relaxation-time parameters defined within the empirical functions employed for data fitting. For the CD function, $\langle\tau\rangle$ can be calculated via $\langle\tau\rangle = \tau_{CD}\,\beta_{CD}$ [56], where $\tau_{CD}$ and $\beta_{CD}$ denote the relaxation time and width parameter within the CD equation [44], respectively. In the present work, we always provide average relaxation times (simply denoted by $\tau$), except for the 14.8 mol% sample from Ref. [26], where the HN function was employed, for which $\langle\tau\rangle$ is undefined [57]. There, we approximated $\langle\tau\rangle$ by the inverse circular loss-peak frequency, calculated from the HN parameters [53].

## IV. RESULTS AND DISCUSSION

### A. Dielectric measurements with moderate cooling rates

Aqueous solutions with salt concentrations $x \gtrsim 10$ mol% can be supercooled with moderate rates (< 1 K/min) not revealing an NML [24,58]. As an example, Fig. 2(a) shows dielectric-loss spectra $\varepsilon''(\nu)$ for $x = 14.8$ mol% as published in Ref. [26]. After subtracting the dc-conductivity contribution, peaks are observed shifting to lower frequencies upon cooling. As shown in [26], with increasing $x$, at room temperature this peak smoothly develops from the structural $\alpha$ relaxation peak of pure water at ~20 GHz (for alternative interpretations of this peak, see [26]). The open circles in Fig. 3 show an Arrhenius plot of $\tau(T)$, estimated from the fits in Fig. 2(a) (see preceding section). Starting at picoseconds in the low-viscosity liquid, it increases by more than 12 decades when approaching the deeply supercooled-liquid regime, thereby crossing pure water's NML (blue-shaded area). The red line in Fig. 3 is a fit of $\tau(T)$ by the commonly employed, empirical Vogel-Fulcher-Tammann (VFT) law [15,16,59]:

$$\tau = \tau_0 \exp\left(\frac{D T_{VF}}{T - T_{VF}}\right) \qquad (1)$$

It reasonably describes the experimental data without any indication of an FST. Similar conclusions were drawn for a 12 mol% solution measured below 3 GHz [24].

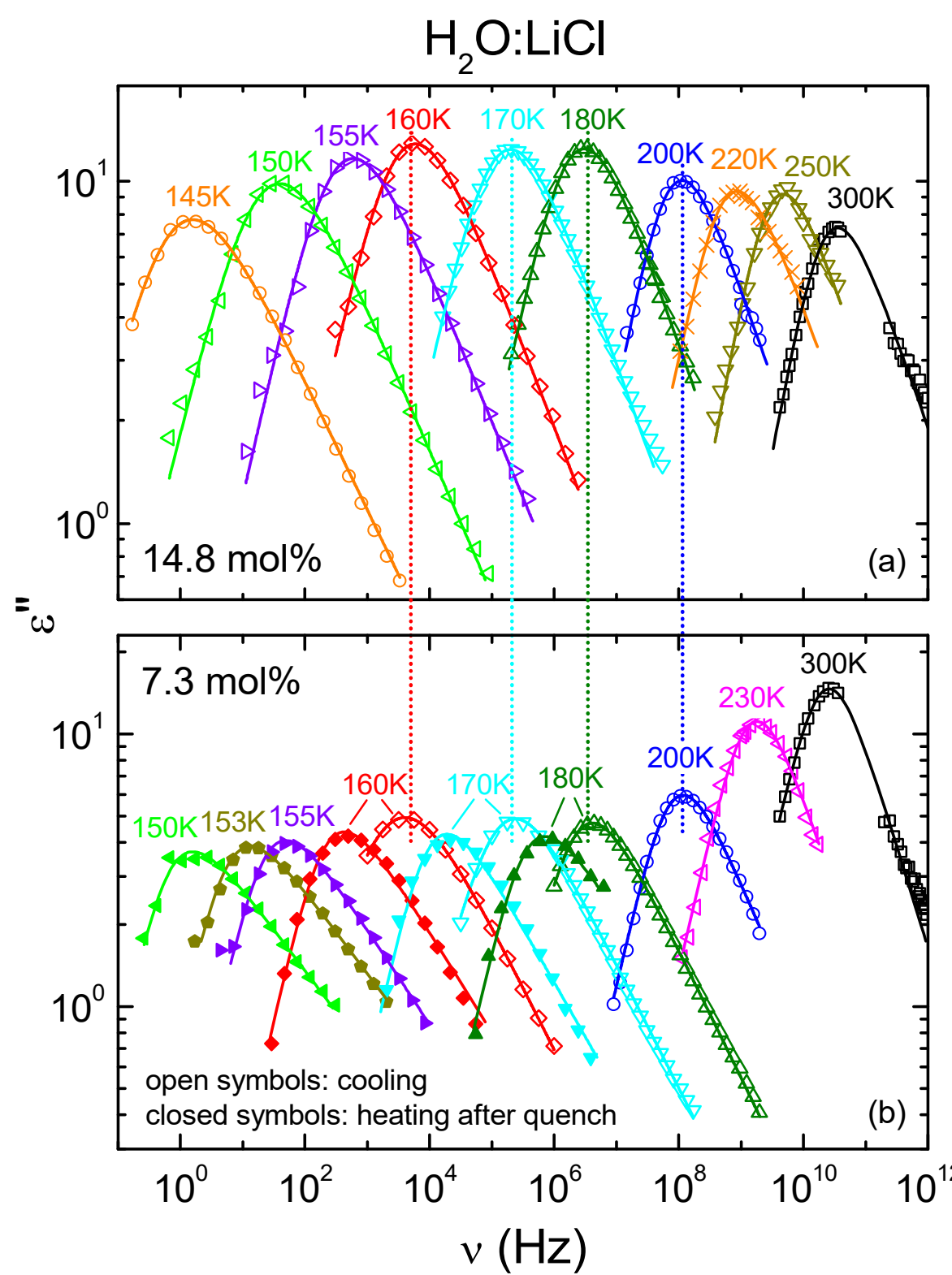

FIG. 2. Dielectric-loss spectra after correction for dc conductivity for (a) a 14.8 mol% solution (from [26]; here we do not show the high-frequency contributions treated in that work) and (b) a 7.3 mol% solution. Open symbols: determined on cooling with ~0.5 K/min. Closed symbols for 7.3 mol%: measured upon heating after quenching in liquid nitrogen. Solid lines in (a) and (b): fits with the empirical HN [52] and CD [44] function, respectively. The vertical lines demonstrate the agreement of the peak positions in the phase-separated 7.3 mol% sample with those of the 14.8 mol% sample.

Such high salt concentrations make it questionable whether the results are of significance for pure water [31]. Hence, we performed measurements at lower concentrations as shown for 7.3 mol% in Fig. 2(b) presenting dc-subtracted data. However, for moderate cooling rates (open symbols), even for 7.3 mol% partial crystallization occurs in the temperature range of the NML. This is evidenced by the finding that the relaxation times derived from fits of the spectra as described in Sec. III agree with the results for 14.8 mol% within experimental resolution (open upright triangles in Fig. 3 at $1000/T > 4$ K$^{-1}$; see also the vertical dotted lines in Fig. 2). For the 1.8 and 3.6 mol% samples, we found similar behavior. Such samples undergo phase separation (see Appendix A for more details) whereupon part of the sample crystallizes into ice (with $x = 0$) upon cooling. Thus, according to the phase diagram from Ref. [60] (Appendix A), the remaining liquid fraction becomes enriched in LiCl (because the salt ions are not incorporated into the ice fraction), until a stable concentration of about 16.7 mol% (pentahydrate; LiCl:5$H_2O$) is reached. Then, the dielectric spectra reflect the dynamics in this pentahydrate fraction and $\tau$ is nearly identical to that of the 14.8 mol% sample. As seen in the inset of Fig. 3, at temperatures above the NML ($1000/T < 4.25$ K$^{-1}$) the $\tau$ data of the low-concentration samples vary continuously between those for pure water (crosses [26] and plusses [61]) and 14.8 mol% (circles). This is reasonable because at these high temperatures no phase separation is expected (see Appendix A).

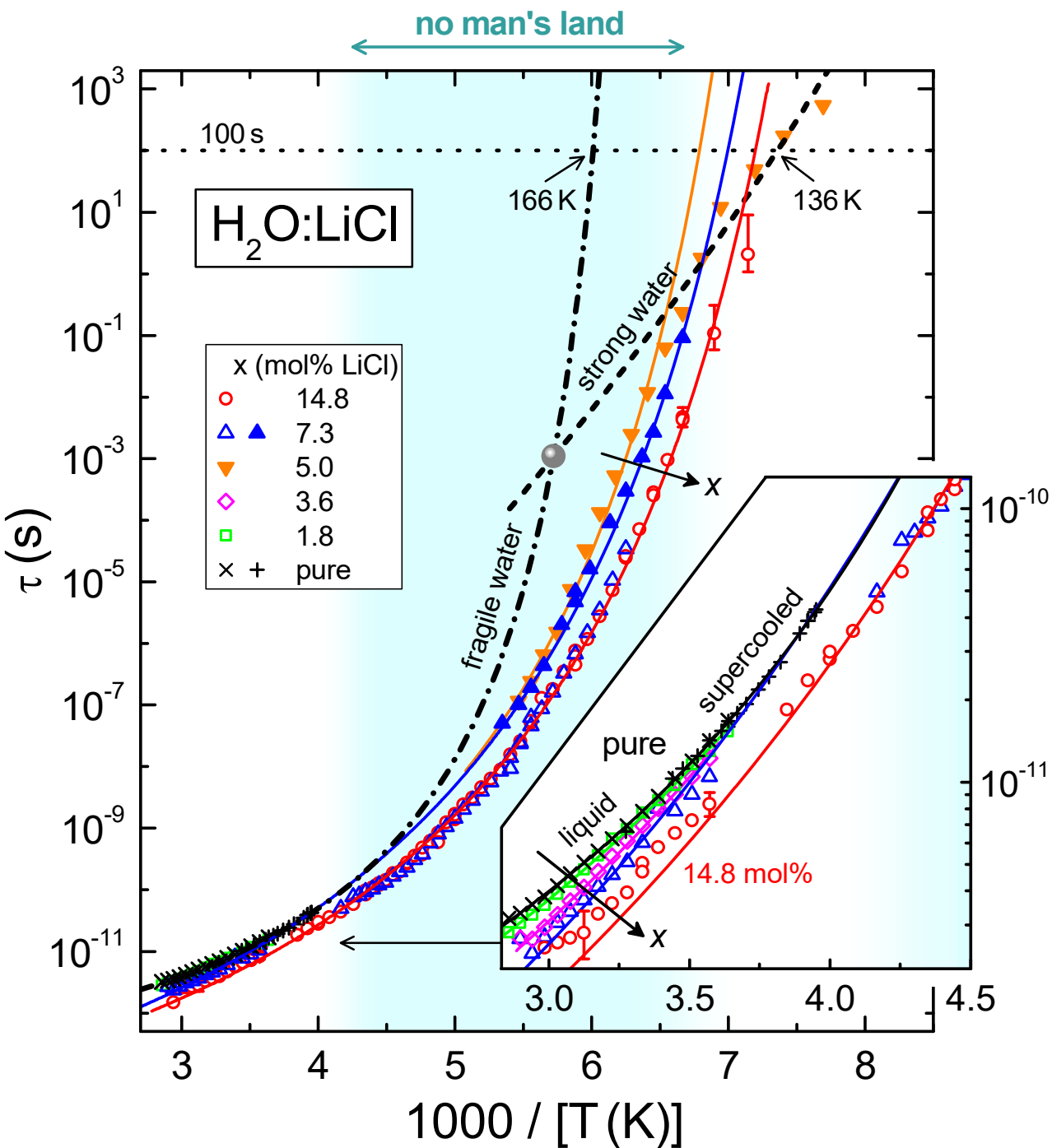

FIG. 3. Temperature dependence of $\tau$ in Arrhenius representation deduced from dielectric spectra for pure water and LiCl solutions. The NML of water is indicated by the shaded region. The inset presents an enlarged view at high temperatures. Open symbols: $\tau$ determined using moderate heating and cooling rates. Closed triangles up and down: quenched samples with 7.3 and 5.0 mol%, respectively. Crosses and plusses: pure water at temperatures above the NML (from [61] and derived from the spectra in [26]). Solid lines: VFT fits [Eq. (1)] for the solutions. Dash-dotted line: VFT fit for liquid and supercooled pure water at high temperatures (crosses and plusses), i.e., for the fragile form of water. Dashed line: estimated VFT law for the suggested strong form of water (see text). Grey sphere: estimated FST of pure water. Dotted horizontal line: $\tau(T_g) = 100$ s.

## B. Phase separation and glass-transition temperatures investigated by DSC

To investigate the glass transitions of the solutions and to confirm the phase-separation scenario, we performed additional DSC measurements. Figure 4(a) shows temperature-dependent DSC scans around the glass transition for aqueous solutions with 7.3, 10.1, 14.8, and 21.7 mol% LiCl. They were carried out with constant heating rates of 10 K/min after precooling with the same rate. The samples with 14.8 and 21.7 mol% LiCl revealed the typical signatures of transitions into homogeneous low-temperature glassy states without any indications of crystallization. As indicated by the dashed lines in Fig. 4(a), from the onset of the step-like increase in the DSC trace, $T_g$ can be determined. In the concentration regime $x < 10$ mol%, the reduced step height indicates partial crystallization. Especially, for the solution with 7.3 mol% LiCl, lying deep within the phase-separation regime (Appendix A), this is corroborated by a complete heating and cooling cycle shown in Fig. 4(b): On cooling, we find crystallization and, subsequently, a weak glass transition, while on heating the glass-transition anomaly is followed by an endothermic melting transition.

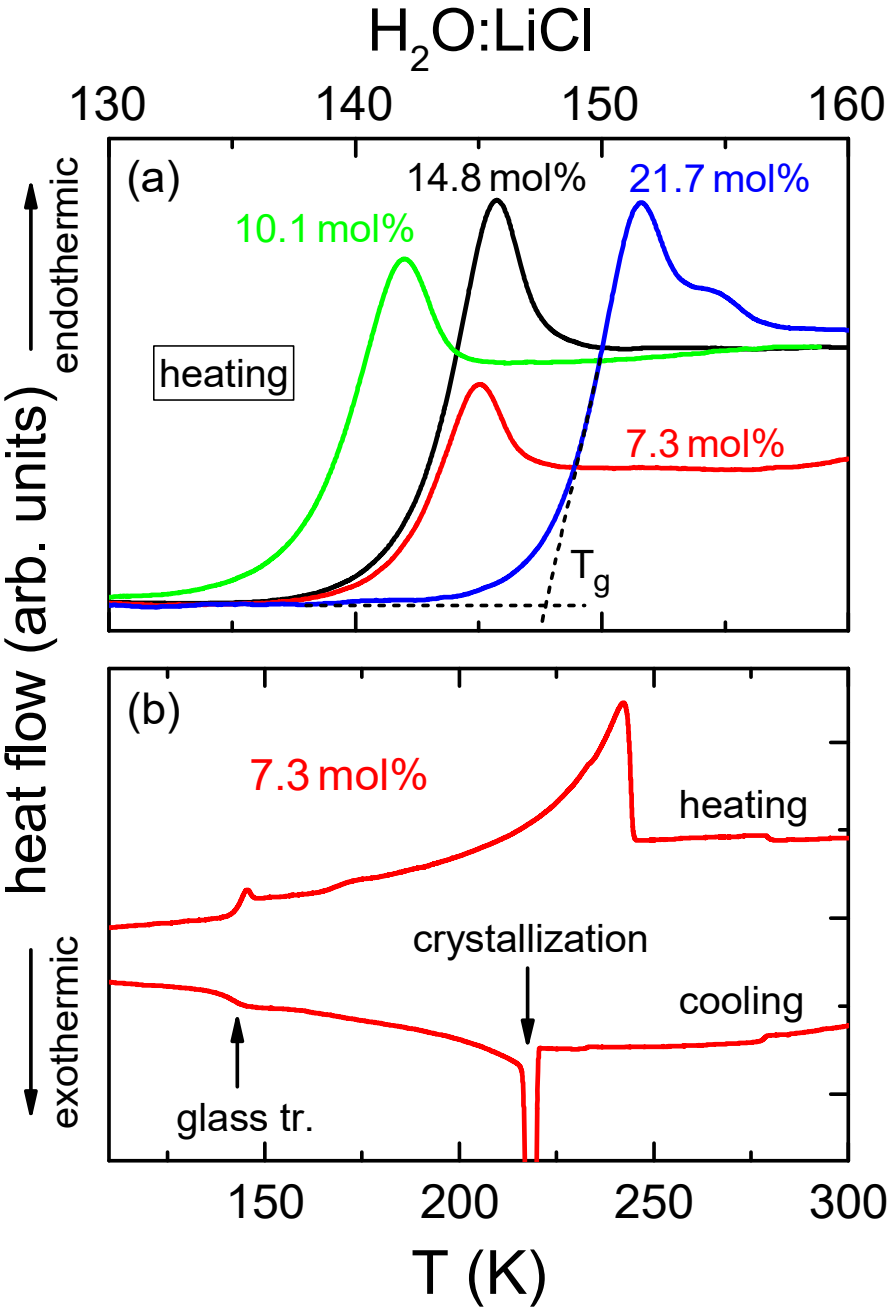

FIG. 4. DSC results for various LiCl concentrations measured with cooling and heating rates of 10 K/min. (a) Glass-transition region of representative samples with 7.3, 10.1, 14.8, and 21.7 mol% measured on heating. The dashed lines for the 21.7 mol% sample indicate the determination of the glass-transition temperature $T_g$. (b) A complete cooling and heating cycle in the full temperature range for an aqueous LiCl solution with 7.3 mol%.

From DSC measurements with different heating rates, temperature-dependent relaxation times can be derived, from which $T_g$ can also be deduced. The method relies on the Kobeko–Frenkel–Reiner relation [62], providing a connection of heating rate and relaxation time:

$$|q|\,\tau(T_g) = C \tag{2}$$

Here, $q$ is the cooling rate, $\tau(T_g)$ the relaxation time at the temperature of the DSC anomaly, and $C$ a constant. Usually, the glass-transition anomaly as observed in DSC measurements with a rate of 10 K/min is attributed to $\tau = 100$ s [63,64,65]. This leads to $C = 16.67$ K, which we also adopt in this work. As an example, Fig. 5(a) shows the application of this method to a 14.8 mol% aqueous LiCl solution. It was cooled with a rate of 10 K/min and subsequently heated with variable heating rates between 2.5 and 30 K/min. The glass-transition temperatures were determined from the onset of the endothermic heat flow [cf. dashed lines in Fig. 4(a)]. The increasing glass-transition temperature with increasing heating rate is clearly visible. In Fig. 5(b), for the 10.1, 14.8, and 21.7 mol% samples, the closed symbols show an Arrhenius representation of the temperature-dependent relaxation times as deduced via Eq. (2) from the rate-dependent DSC measurements discussed above. They span about two decades in $\tau$ around the calorimetric glass transition. These data enable the determination of the glass-transition temperature by applying the often-used condition $\tau(T_g) \approx 100$ s, thereby supplementing the direct determination of $T_g$ from the DSC traces as indicated in Fig. 4(a).

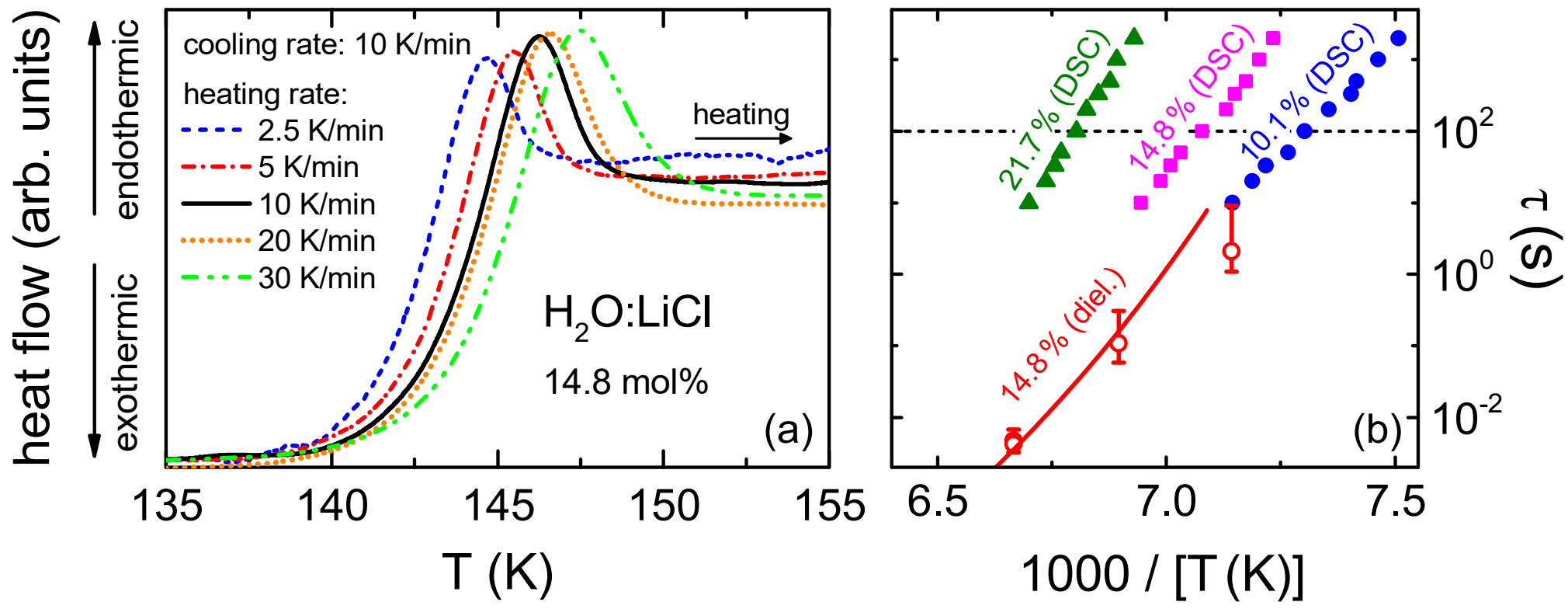

FIG. 5. (a) Measurements of a 14.8 mol% LiCl solution with variable heating rates after cooling with 10 K/min. (b) The closed symbols show Arrhenius plots of $\tau(T)$, derived from rate-dependent DSC measurements as presented in frame (a), using Eq. (2) with $C = 16.67$ K, for three salt concentrations. For comparison, the open circles show results as deduced from dielectric spectroscopy in the 14.8 mol% sample in this temperature range. The solid line is a VFT fit [Eq. (1)] of the complete $\tau(T)$ trace of this sample as shown in Fig. 3. The dashed line indicates $\tau(T_g) = 100$ s.

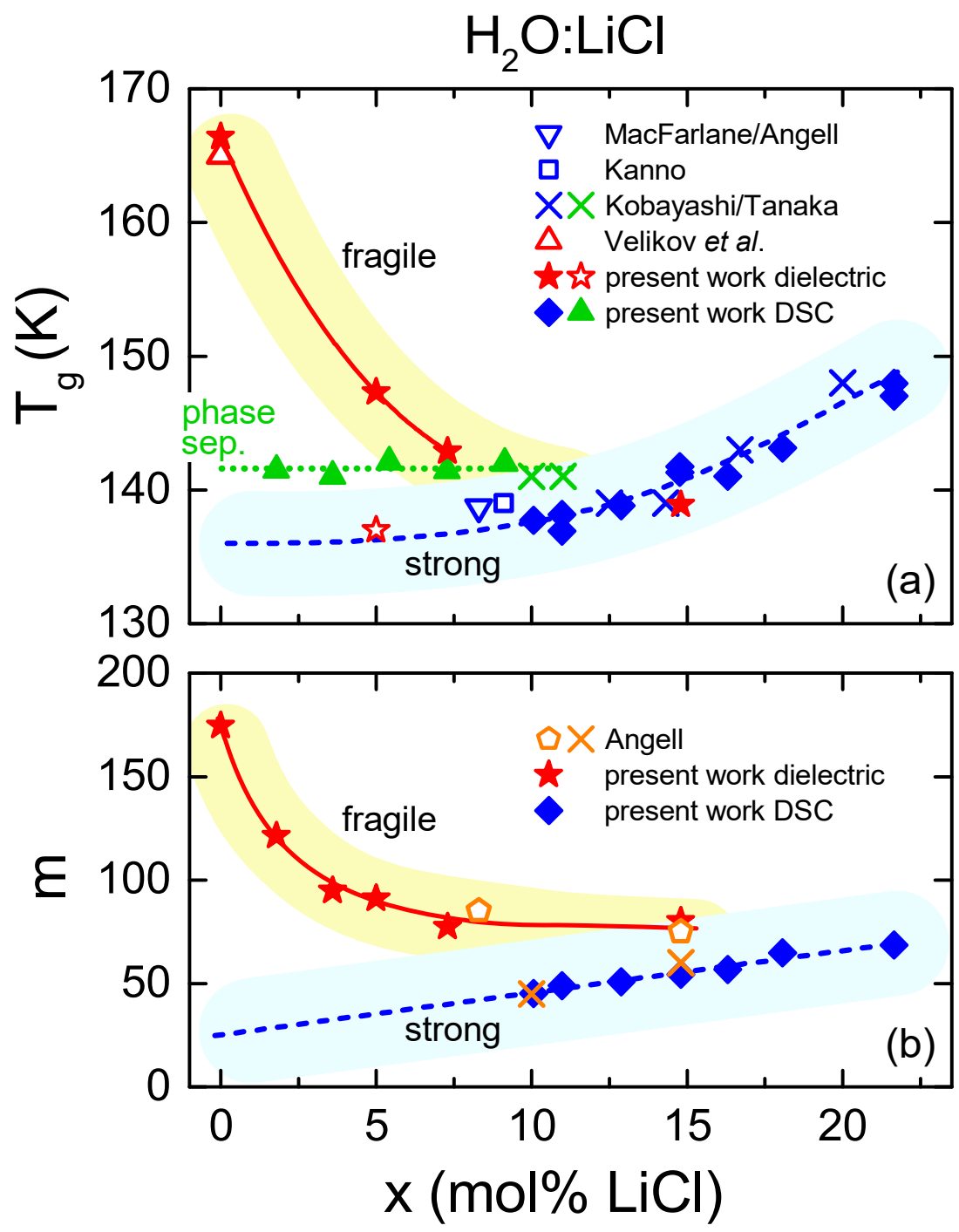

FIG. 6. (a) Glass-transition temperatures $T_g$ and (b) fragility indices $m$ in dependence of $x$ in the two forms of water (fragile: yellow shaded; strong: blue shaded). All lines are drawn to guide the eye. Closed diamonds and closed triangles: present results from DSC experiments (Figs. 4 and 5). The $T_g(x)$ results in (a) are complemented by literature data from MacFarlane and Angell [68], Kanno [67], and Kobayashi and Tanaka [66] as indicated in the legend. Green symbols: results for phase-separated samples. Blue symbols: homogeneous samples investigated around $T_g$, in the supposed strong water state. Stars: results derived from the dielectric $\tau(T)$ (Fig. 3) of homogeneous samples, including pure water and the quenched samples at 5 and 7.3 mol%. Upright open triangle in (a): $T_g$ of pure water, proposed in [78]. Pentagons and crosses in (b): $m(x)$ for the fragile and strong states of LiCl solutions, respectively, published by Angell [6].

For comparison, in Fig. 5(b) we included the results for 14.8 mol% from dielectric spectroscopy as documented in the full temperature range in Fig. 3. We find that the slope of $\tau(1/T)$ observed in the dielectric results comes close to that found in the DSC experiments. However, there seems to be a small but significant mismatch of the absolute values of $\tau$, which may require a redefinition of the constant $C$ defined in Eq. (2). In Fig. 6(a), this small mismatch is mirrored by the difference between the red star and the blue diamond at 14.8 mol%, showing $T_g$ from the dielectric and DSC measurements, respectively. However, as this difference is within the data scatter of the DSC data in Fig. 3(a), we consider this issue as unimportant, and it does not affect the conclusions of the present work.

The closed triangles and diamonds in Fig. 6(a) show $T_g(x)$ deduced from the DSC measurements. For the 10.1, 14.8, and 21.7 mol% solutions, results from both methods discussed above are provided, which explains the presence of two slightly different values. For $x \geq 10$ mol% (diamonds), $T_g(x)$ exhibits an increase with increasing LiCl content, in agreement with earlier results [20,66,67,68]. However, for lower concentrations (triangles), $T_g$ is essentially constant. Its value of ~142 K, consistent with literature data for 10 and 11 mol% (green crosses at 10 and 11 mol%) [66], corresponds to that of a homogeneous solution of about 16–17 mol% and reflects the liquid pentahydrate fraction of phase-separated samples. Interestingly, the glass-transition temperatures at $x > 10\%$ can be reasonably extrapolated to a $T_g$ of 136 K at $x = 0$ (dashed line). This lies within the range often assumed for pure water [39,69,70,71,72] [however, see [21] for a more complex $T_g(x)$ behavior of hyperquenched solutions, nevertheless leading to a similar $T_g(x \rightarrow 0)$].

### C. Dielectric measurements of quenched samples

While low-concentration LiCl solutions partially crystallize for moderate cooling rates, for them the inaccessibility of the NML may be less strict than for pure water. Indeed, at least for the samples with 5 and 7.3 mol%, we managed to avoid any crystallization by quenching them in liquid nitrogen. Typical conductivity-corrected spectra, measured upon heating after the quench, are shown for 7.3 mol% in Fig. 2(b). For 160, 170, and 180 K, the quenched (closed symbols) and the phase-separated samples (open symbols) reveal significantly different peak positions. Moreover, in contrast to the latter (cf. dotted lines in Fig. 2), for the quenched samples the peak frequencies are not similar to those of the 14.8 mol% sample [Fig. 2(a)], speaking against phase separation. Fitting the spectra of both quenched solutions (for details, see section III]) leads to $\tau(T)$ as shown by the closed upright and inverted triangles in Fig. 3. Within the NML, $\tau$ of these quenched solutions increases with decreasing salt content, confirming that phase separation plays no role here. This systematic variation (also including the homogeneous 14.8 mol% sample) indicates that the relaxation time of pure water within the NML should be even higher. The lines through these data points are fits with the VFT formula, Eq. (1) [73]. The mentioned increase of $\tau$ with decreasing $x$ essentially can be traced back to an increase of $T_g$, as estimated from the condition $\tau(T_g) \approx 100$ s [stars in Fig. 6(a) for $x = 14.8$, 7.3, and 5%].

Figure 2(b) reveals that the peak amplitudes for the unquenched (phase-separated) and the quenched solutions are of similar order. However, one should be aware that the latter was violently quenched in liquid nitrogen and the corresponding sudden thermal contraction of sample and capacitor may have affected the proper detection of absolute values of the permittivity, e.g., due to the formation of cracks or an incompletely filled capacitor.

### D. Glass-transition temperature and fragility of pure water above the NML

The relaxation-time data for pure water in Fig. 3 (crosses and plusses), measured above the NML [26,61], are shown in more detail in Fig. 7. The crosses (×) were determined in the present work by analyzing the spectra previously reported by Lunkenheimer *et al.* [26], covering about 100 MHz – 20 THz. Notably, in contrast to previous publications, these spectra include the real part of the permittivity and both the low- and high-frequency flanks of the loss peaks. This leads to unprecedented precision of the derived $\tau(T)$ data without having to resort to any assumptions about the spectral shapes or amplitudes of the relaxation features. The $\tau(T)$ data as reported by Bertolini *et al.* [69] (plusses; also shown in Fig. 3) nicely extend those deduced from Ref. [26] to lower temperatures. These two data sets agree almost perfectly with those reported in smaller temperature ranges by Kaatze [74] (closed squares) and by Buchner *et al.* [75] (open circles), also included in Fig. 7. Only those by Rønne *et al.* [76,77] (closed diamonds) somewhat deviate, especially at the lower temperatures, suggesting weaker deviations from Arrhenius behavior. Thus, we think these data should not be employed for an estimation of the fragility or $T_g$ of pure water at high temperatures.

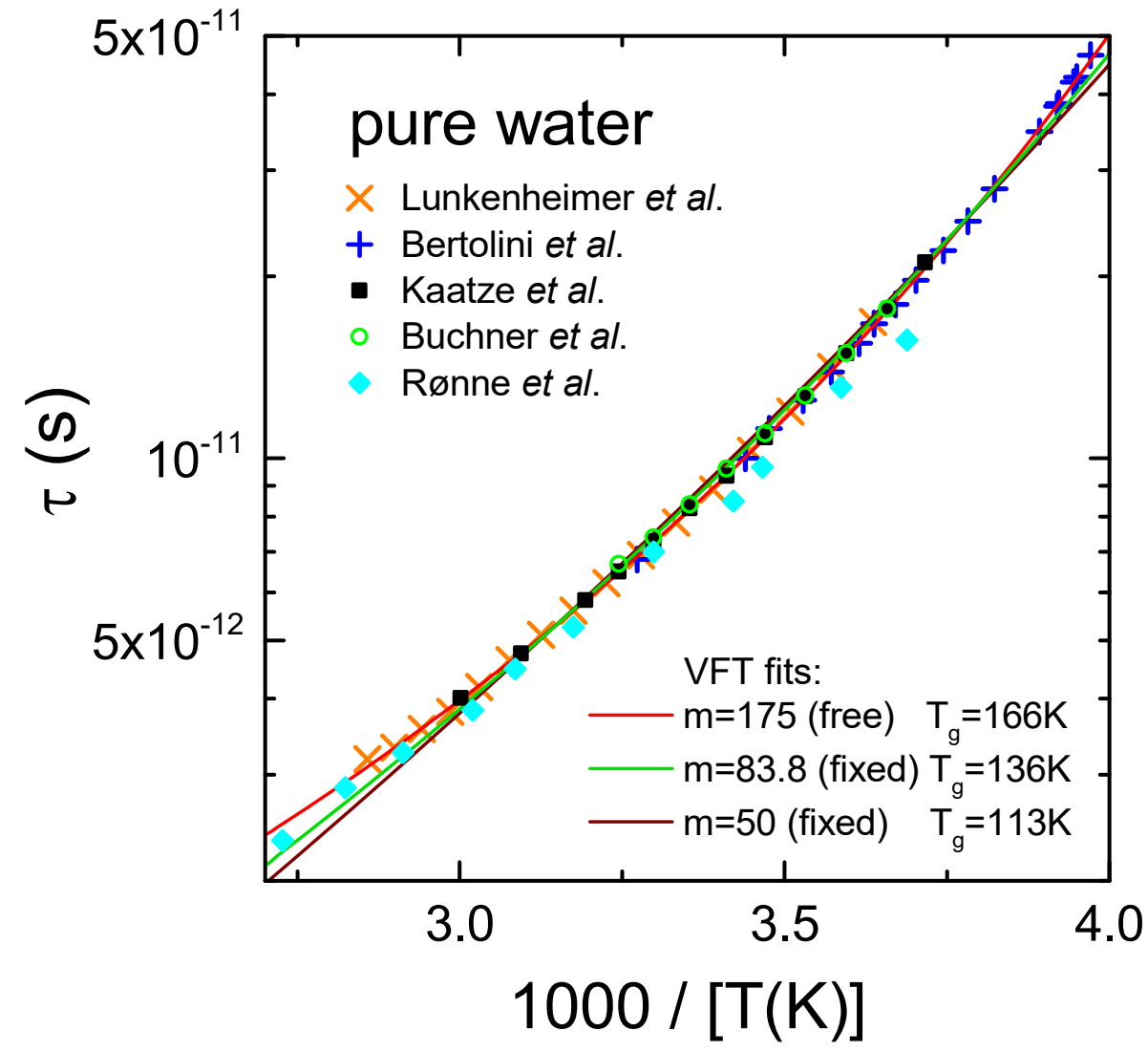

FIG. 7. Arrhenius representation of $\tau(T)$ of pure water at temperatures above the NML derived from data by Lunkenheimer *et al.* [26] (crosses) and taken from Bertolini *et al.* [61] (plusses) as also shown in Fig. 3. The lines are VFT fits [Eq. (1)] with different (partly fixed) fragility indices $m$ as indicated in the lower legend (see [59] for the relation of $m$ and $D$). The free fit curve (red, upper line) is the same as the dash-dotted line in Fig. 3. For comparison, also the data from Kaatze *et al.* [74], Buchner *et al.* [75], and Rønne *et al.* [76,77] are included (see upper legend).

The red, upper line in Fig. 7 is a fit of the combined data as derived from Ref. [26] and taken from Ref. [61], using the VFT formula, Eq. (1). Leaving all parameters free leads to $T_g = 166$ K [determined via $\tau(T_g) \approx 100$ s]. From the strength parameter $D$ in the VFT equation [Eq. (1)], the fragility index $m$, used to quantify the deviations from Arrhenius behavior, can be deduced [59]. We obtain a very high fragility of $m = 175$. To check for the significance of these parameters, the green, middle line in Fig. 7 shows a fit with $m$ fixed to 83.8, which is adjusted to lead to $T_g = 136$ K, a value often assumed for water [39,70,71,72]. The deviations of fit and experimental data, however, are significant, although $m$ still is rather large. The brown, lower line, where $m$ was fixed at an even smaller value of 50 (intermediate between strong and fragile) demonstrates that the experimental data on pure water above the NML are clearly incompatible with strong dynamics where $m$ would be close to its minimum value of 16 [59]. Obviously, $\tau(T)$ of pure water cannot be reasonably fitted with $m$ below about 100, and it is clearly incompatible with $T_g = 136$ K, except when assuming a transition to weaker temperature dependence at lower temperatures. As discussed in Appendix B, the $T_g$ of 166 K derived from the free fit in Fig. 7 has an uncertainty of about ±15 K as it is essentially based on an extrapolation of the VFT fit of the low-viscosity-liquid data up to 100 s. Nevertheless, it reasonably matches with the $T_g(x)$ trend revealed by the low-$x$ solutions [closed stars in Fig. 6(a)] and with $T_g$ values reported for water above the NML [78,79]. It also is approximately consistent with the empirically founded 2/3 rule, stating that $T_g/T_m \approx 2/3$ (where $T_m$ is the melting temperature) [16,80,81], which leads to $T_g \approx 182$ K.

As mentioned above, the fragility index $m \approx 175$ deduced from the present high-precision data of pure water is extremely high. High fragility of water above the NML was also reported based on thermodynamic measurements [5,78] and on viscosity data, covering a similar temperature range as the pure-water $\tau(T)$ data in Fig. 3 [82]. The obtained fragility indices of pure water and of the homogeneous solutions are shown by the stars in Fig. 6(b). Here we also include data for 1.8 and 3.6 mol%, for which experimental data in the homogeneous state are only available in a rather small temperature range at high temperatures (inset of Fig. 3). Thus, we have fitted them using fixed values for $T_{VF}$, obtained by interpolation. A continuous trend of $m(x)$ of the solutions, deduced from the dielectric data, is revealed, well consistent with the obtained high fragility of pure water.

### E. Fragility determined by DSC

We also performed DSC measurements with different heating/cooling rates to determine the fragility according to Ref. [43]. This can be done by evaluating the obtained $\tau(T)$ data as shown in Fig. 5(b) within an Angell plot [83], log($\tau$) vs. $T_g/T$ [59]. A well-established alternative method to more directly determine a liquid's fragility via calorimetric measurements is the so-called enthalpy-relaxation method developed by Angell and co-workers [43,84] as well as Yue *et al*. [85]. It can reliably reproduce fragility values obtained from viscosity or relaxation-time measurements [43,86]. Figure 8 demonstrates the application of this method to the 14.8 mol% solution. Here, the cooling rates were varied between 0.5 and 100 K/min and, subsequently, the sample was heated up with the standard scanning rate of 10 K/min. As expected for this method [43], the overshoot of the endothermic peak at the glass transition, measured during heating, strongly depends on the preceding cooling rate. As explained in Ref. [43], the cooling-rate dependent enthalpy release $\Delta H(q)$ can be determined by comparing the scans with variable cooling rates to the standard scan. The fictive temperatures $T_f$ corresponding to the different cooling rates can be calculated using $T_f = T_f^s + \Delta H(Q)/\Delta c_p$ [43]. Here $T_f^s$ is the fictive temperature, determined from a standard scan using the method described by Moynihan *et al*. [87] and $\Delta c_p$ is the difference in heat capacity between glass and supercooled liquid. Following this procedure, $m$ can be directly estimated from the slope of the reduced cooling rate log($q/q_s$) versus the inverse reduced fictive temperature $T_f^s/T_f$.

We found an almost perfect agreement of the fragility values as determined from both DSC-based methods described above. The obtained concentration dependence of $m$ is shown by the closed diamonds in Fig. 6(b). It reveals an approximately linear decrease with decreasing concentration. Such a decrease was also reported in [66] where, however, different $m$ values were found for different experimental methods.

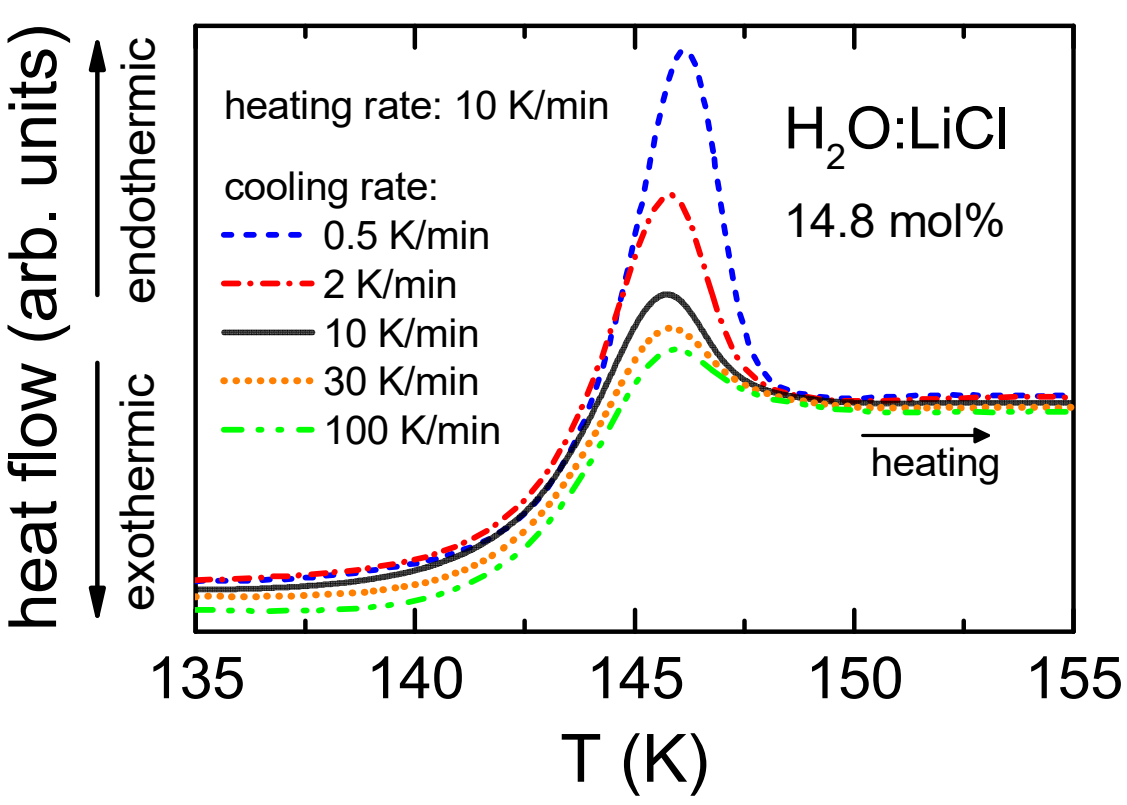

FIG. 8. Application of the enthalpy relaxation method [43,84,85]: The glass transition as revealed during heating with 10 K/min, after cooling the glass with different rates as indicated in the legend.

### F. Implications for the FST of pure water

An extrapolation of $m(x)$ derived from the DSC measurements as discussed in the preceding section [dashed line in Fig. 6(b)] leads to $m \approx 25$ for pure water, consistent with the notion that deeply supercooled water is a strong liquid [4,28,72]. Using $m = 25$ and $T_g = 136$ K from the above-discussed extrapolations (dashed lines in Fig. 6), and assuming $\tau_0 = 10^{-14}$ s as a standard value for the inverse attempt frequency [59], enables a rough estimate for $\tau(T)$ of water close to $T_g$, where the DSC experiments were performed (dashed line in Fig. 3). It clearly does not match the higher-temperature experimental data of liquid and supercooled water without assuming an anomaly. Similar conclusions from data above and below the NML were drawn previously [4,5,10,22,42].

Based on the crossing of the dashed and dash-dotted lines in Fig. 3, the sphere at 175 K represents a rough estimate of the FST temperature of water (the actual crossover probably is more smeared out [6,42]). In literature, values between 190 and 228 K were reported [5,8,12,17,29]. The $\tau(T)$ traces of fragile and strong pure water within the NML, indicated in Fig. 3, should be regarded as one possible scenario only. Nevertheless, one should note that the dash-dotted VFT line in Fig. 3 is based on $\tau(T)$ data of unprecedented precision as discussed above. Even if one dismisses the roughly estimated dashed line, this VFT fit (or corresponding ones with lower or higher $T_g$ within its uncertainty of ±15 K; see Appendix B]) does not smoothly match the Arrhenius behavior of deeply supercooled liquid water, reported below the NML [1,10,72]. Therefore, even our pure-water data alone indicate that somewhere within the NML an FST must occur. Based on these data alone, we cannot say whether $\tau(1/T)$ exhibits an abrupt or more smeared-out change of slope. However, as discussed below, in the quenched 5 mol% LiCl solution, the latter is found.

Figure 6 reveals that, for the homogeneous low-concentration LiCl solutions, both $T_g$ and $m$ markedly increase with decreasing salt content (closed stars). This strongly contrasts with the *decrease* of both quantities found for the high-concentration solutions (closed diamonds). For the fragility, these opposite trends were already pointed out by Angell [6], based on sparse viscosity and DSC data [pentagons and crosses in Fig. 6(b)], and interpreted as indicative of an FST of pure water. The more comprehensive data of the present work nicely confirm this view. For 14.8 mol%, the values of $T_g$ from the dielectric and DSC experiments are in accord and $m$ does not differ dramatically. In contrast, $m(x)$ and $T_g(x)$ at lower concentrations clearly do not agree with the extrapolated behavior of the samples with $x > 10$ mol% (dashed lines in Fig. 6). That is, an extrapolation to $x = 0$ of the dielectric (or viscosity [6]) data leads to markedly different properties of pure water than an extrapolation of the DSC results. To understand this apparent discrepancy, one should note that the latter are based on low-temperature measurements, around $T_g$. In contrast, the analysis of the dielectric experiments is based on results above $T_g$, mostly extending far into the low-viscosity liquid range. As for 14.8 mol% both methods lead to comparable results, one can conclude that, at high concentrations, there is essentially only a single state of these solutions, persisting at all temperatures [24]. The extrapolation of its properties to pure water is in accord with the strong state of pure water with $T_g \approx 136$ K as deduced from low-temperature data on hyperquenched or pressure-driven amorphous water [69,70,71,72]. However, for the low-concentration solutions and pure water, an additional fragile state exists at high temperatures. Its $T_g(x)$ and $m(x)$ significantly increase for decreasing salt contents. As indicated by the solid lines in Fig. 6, this increase is in full accord with $T_g \approx 166$ K and with the high fragility of 175 obtained from the VFT fit of pure water above the NML.

### G. Direct detection of the FST in the quenched 5 mol% solution

If indeed two states of water exist, a transition (not necessarily a phase transition) between them should occur upon temperature variation, which, however, is unobservable due to the NML. Can it be detected in the quenched solutions that allow exploring this region? As seen in Fig. 3, just as for the 14.8 mol% sample (circles), $\tau(1/T)$ for 7.3 mol% LiCl (closed upright triangles) does not exhibit any anomaly up to the highest $\tau$ values of about 0.1 s, corresponding to loss-peak frequencies of the order of 1 Hz. However, for the 5 mol% solution we have extended the measurements down to 100 μHz (Fig. 1). This enables the direct detection of a crossover of $\tau(1/T)$ (inverted triangles in Fig. 3) into a significantly weaker temperature variation at low temperatures, characteristic of a stronger liquid. Using $\tau(T_g) \approx 100$ s, we arrive at $T_g \approx 137$ K for this strong state of the solution [open star in Fig. 6(a)], in good accord with the extrapolated $T_g(x)$ curve of the strong high-$x$ solutions (dashed line).

## V. SUMMARY AND CONCLUSIONS

In summary, we have obtained strong hints at two forms of pure water at high and low temperature, separated by a rather smooth FST. Our conclusions are based on the following achievements: (i) The unprecedented precision of our $\tau(T)$ data for pure water at high temperatures, proving its high fragility and excluding a match to low-temperature data of water without an FST. (ii) The direct detection of a smooth FST for a quenched low-concentration sample, to our knowledge never achieved before in an aqueous solution. (iii) The contradicting $T_g(x)$ and $m(x)$ behavior detected by broadband dielectric spectroscopy at high and by DSC at low temperatures. The emergence of this phenomenon at low concentrations, while it is absent at high ones, makes it implausible that it should be absent in pure water.

Our results solve the controversial question regarding the occurrence of fragile water and the FST. Moreover, they offer a solution for the controversy on $T_g$ of water: The commonly accepted $T_g$ of 136 K refers to the strong form of water which persists at low temperatures and, thus, is water's true glass-transition temperature. The value of 166 K characterizes the fragile supercooled-liquid and liquid forms of water at high temperatures. However, below a temperature $T_{FST} > 166$ K (roughly estimated here to be ~175 K), it undergoes a smooth FST before glassy freezing at 166 K can occur, shifting the glass transition to 136 K. Notably, a $T_g$ of 136 K is only compatible with the high-temperature water data when assuming an FST, i.e., $T_g \approx 136$ K implies an FST and vice versa. These considerations are irrespective of the microscopic origin of the FST, for which various explanations were proposed [8,10,11,42,88].

## ACKNOWLEDGMENTS

We thank M. Schwab for performing part of the dielectric measurements beyond GHz.

## APPENDIX A: PHASE DIAGRAM AND PHASE SEPARATION OF $H_2O$:LiCl

Detailed binary equilibrium phase diagrams of $H_2O$:LiCl were reported in Refs. [60,89,90]. They exhibit a eutectic point at a temperature between 190 and 200 K and at a LiCl concentration of 8 mol/kg($H_2O$), corresponding to approximately 12.6 mol% LiCl (Fig. 9 [60]). On further increasing salt concentrations, the melting line increases again and exhibits a sequence of peritectic points. In addition to the pure phases, four solid lithium-chloride hydrates, with 1, 2, 3, and 5 water molecules, respectively, were identified. In the dilute aqueous limit, the most stable configuration is the pentahydrate LiCl:5$H_2O$ (Li5), which corresponds to a concentration of $x = 16.7$ mol%. At lower LiCl concentrations and low temperatures, the pentahydrate coexists with pure hexagonal ice (Li5 + ice) and undergoes complete phase separation, while at higher concentration regimes of coexisting hydrate crystals (e.g., Li5 + Li3) were identified.

As documented in the course of this work and as reported in literature, in the LiCl:$H_2O$ system, by conventional cooling rates of bulk systems, glassy low-temperature states can be observed for molar salt concentrations $10 < x < 25$ mol%.

Part of our results from DSC measurements are included in Fig. 9. The squares show the obtained glass-transition temperatures as also included in Fig. 6(a). Upon cooling with moderate rates of the order 0.5 K/min, solutions with $x < 10$ mol% undergo crystallization of only a fraction of the sample close to the melting line of the phase diagram (closed circles in Fig. 9). This results in a phase-separated state of crystalline water and Li5. As explained in Sec. IV A, therefore the constant $T_g \approx 142$ K in this $x$ regime reflects the glass transition of the remaining liquid Li5 fraction. Somewhat below the eutectic point $E$, recrystallization was detected on heating (triangles), which was subsequently followed by an endothermic melting transition. Recrystallization on heating was previously reported for LiCl-doped water samples with $x < 10$ mol% [21,67].

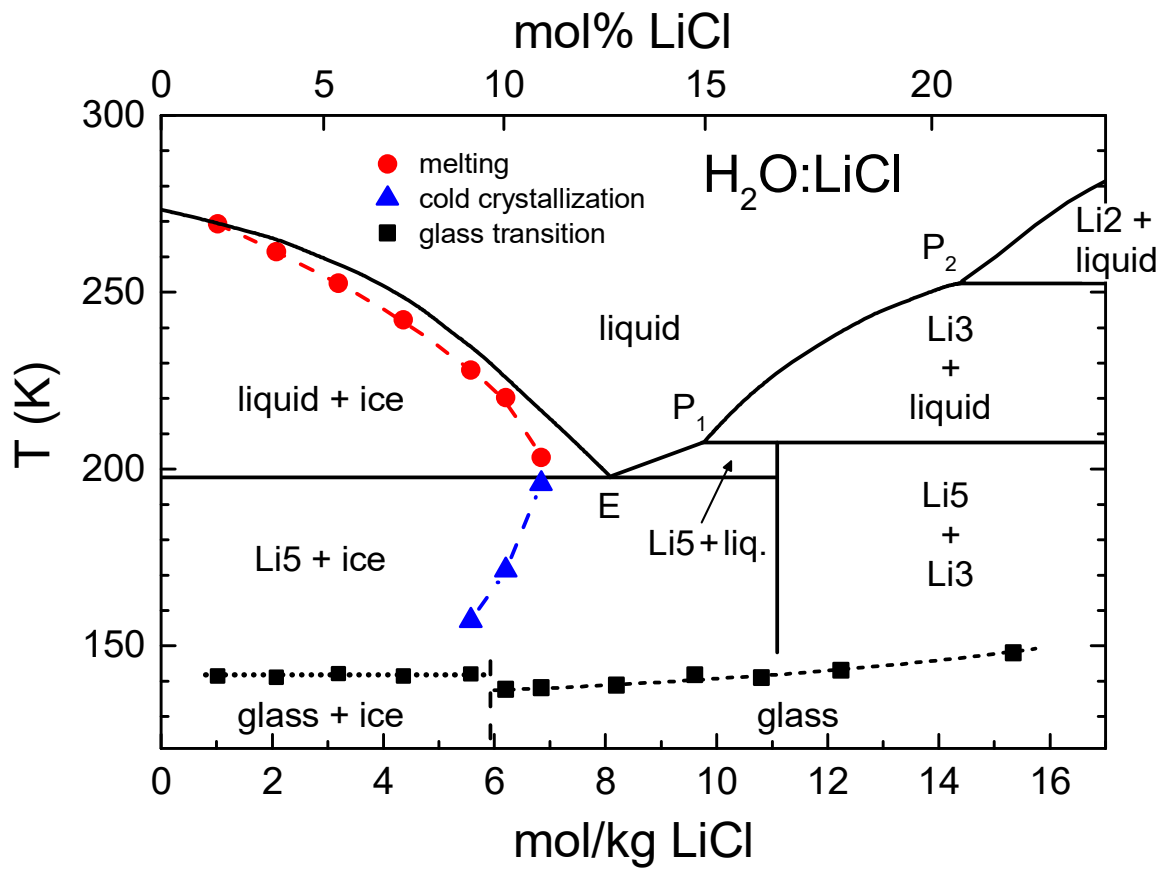

FIG. 9. Low-concentration equilibrium ($x$,$T$) phase diagram of $H_2O$:LiCl as determined by Monnin *et al*. [60]. Solid lines indicate equilibrium phase boundaries. Li2, Li3, and Li5 stand for phases with 2, 3, and 5 water molecules per LiCl, respectively. E indicates the eutectic point. $P_1$ and $P_2$ denote peritectic points. The symbols show DSC results from the present work. Closed Circles: crystallization observed upon cooling. Squares: glass-transition temperatures (depending on the cooling rate, below about 10 mol% a glassy LiCl:5$H_2O$ phase coexists with crystalline water, see text). Triangles: recrystallization temperatures of the supercooled liquid fraction on heating. Dotted and dashed lines are drawn to guide the eye.

## APPENDIX B: UNCERTAINTY OF $T_g$ OF PURE WATER ABOVE THE NML

In Fig. 10(a), we compare the $\tau(T)$ data for pure water at temperatures above the NML, derived from Refs. [26] (×) and [61] (+), to those for deeply supercooled liquid water (termed "low-density liquid"), reported at temperatures below the NML by Amann-Winkel *et al*. (stars) [72]. The dash-dotted line shows the free VFT fit [Eq. (1)] of these data (same as the dash-dotted line in Fig. 3 and as the red solid line in Fig. 7) leading to $T_g \approx 166$ K. Using the commonly assumed condition $\tau(T_g) \approx 100$ s, from Eq. (1) one can derive:

$$T_{\mathrm{VF}} = T_{\mathrm{g}} \left[1 + \frac{D}{\ln(100\mathrm{s}/\tau_0)}\right]^{-1} \tag{B1}$$

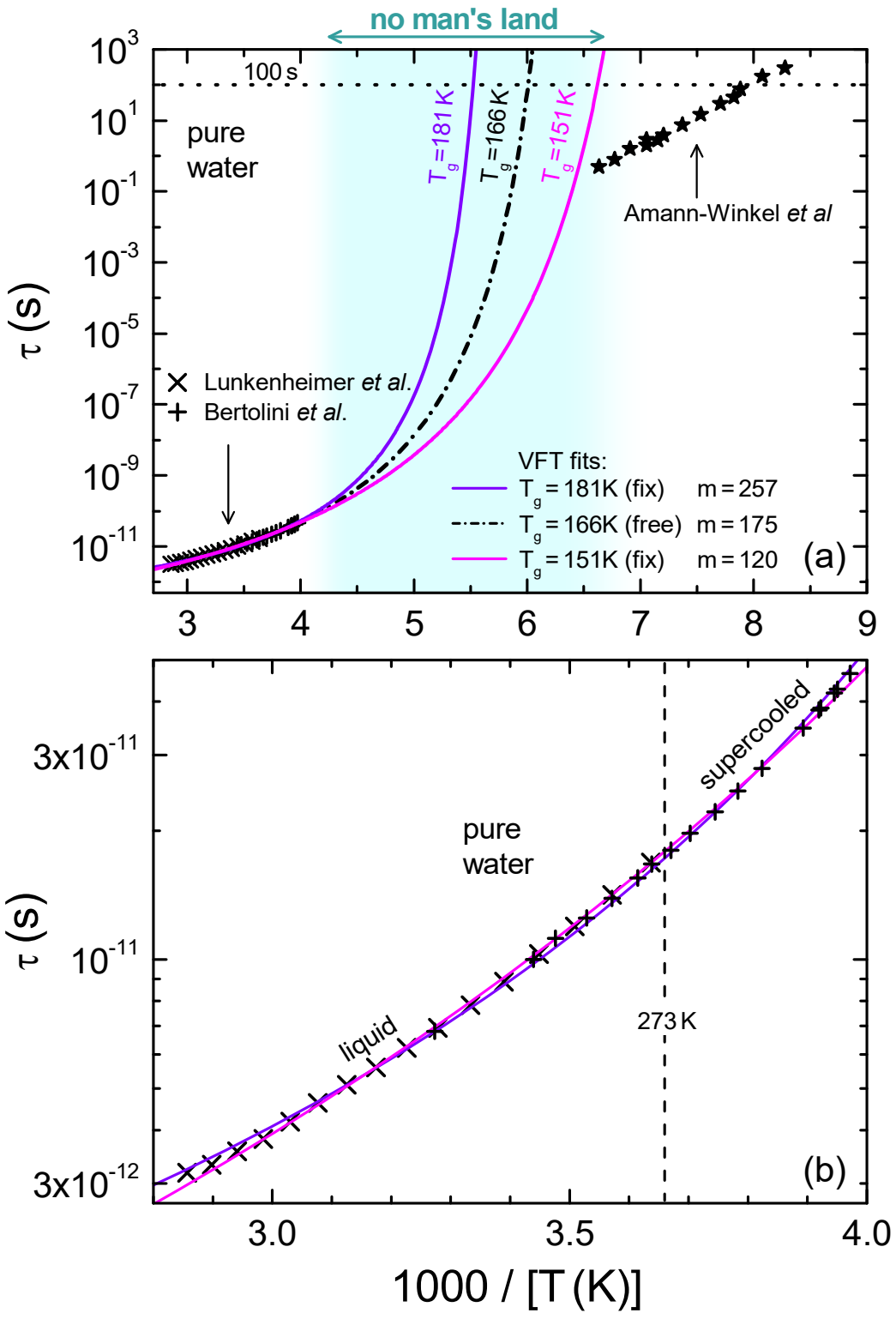

FIG. 10. Relaxation time of pure water above and below the NML (Arrhenius plot). (a) Crosses and plusses: $\tau(T)$ above the NML (shaded region) from Ref. [61] and based on Ref. [26], respectively, as also shown in Figs. 3 and 7. The closed stars show data for deeply supercooled water below the NML from Ref. [72]. The lines are three alternative VFT fits [combining Eqs. (1) and (B1)] of the high-temperature data with free parameters and with $T_g$ fixed to values that are 15 K lower or higher that 166 K (see legend). (The dash-dotted line is the same as the dash-dotted line in Fig. 3 and as the red solid line in Fig. 7.) (b) Enlarged view of the high-temperature data, including the two fits with highest and lowest $T_g$.

This allows using $T_g$ instead of $T_{VF}$ as a parameter in the VFT fits. Figure 10(b) shows two VFT fits of the high-temperature water data with $T_g$ fixed to values that are 15 K higher (violet line) or lower (magenta line) than $T_g \approx 166$ K obtained by the free fit. 15 K can be regarded as an estimate of the uncertainty of $T_g$ as the two alternative fits in this figure just meet the experimental data points, whose size (at least for those determined from our own spectra [26]) approximately corresponds to their uncertainty (about 10%). The same two alternative fit curves are also included in Fig. 10(a). None of the three VFT fits shown smoothly matches the experimental results on deeply supercooled water from Ref. [72]. As pointed out in Sec. IV F, this implies that there is an FST.

Data on extremely supercooled water may be criticized because the methods used to produce it could lead to different types of water, compared to water that hypothetically would be reached by slow supercooling if crystallization would not intervene [31]. However, as discussed in Sec. IV F, the above conclusions on the inevitable occurrence of a FST can also be drawn when estimating the low-temperature behavior of pure water by the extrapolation-based dashed line in Fig. 3.

---